\documentclass[12pt]{iopart}
\usepackage{iopams}
\newcommand{\be}{\begin{equation}}
\newcommand{\ee}{\end{equation}}

\begin{document}

\rapid[de Broglie--Bohm mechanics]{Density Relaxation in de Broglie--Bohm Mechanics}

\author{A F Bennett}

\address{College of Oceanic and Atmospheric Sciences, Oregon State University, Corvallis, OR 97331-5503, USA}

\ead{bennett@coas.oregonstate.edu}

\begin{abstract}

A Lagrangian analysis explains the numerical results of Valentini and Westman (2005) which demonstrate that an initially arbitrary particle density, stirred by the field of de Broglie velocities associated with a Schr\"{o}dinger wave function, relaxes to the Born probability density provided both densities are coarse--grained. 
\end{abstract}

\submitto{\JPA}
\maketitle

\section{Introduction}
The de Broglie--Bohm interpretation of quantum mechanics (Bell, 1987; Bohm and Hiley, 1993; Holland, 1993) continues to attract critical appraisal (e.g., Passon, 2006) in spite of being beset by several difficulties. The most immediately apparent is that the initial density for ontological particles may not be freely chosen, but must be the Born probability density at that time if the two densities are to agree subsequently. The resolution of the difficulty is indicated by  numerical calculations of Valentini and Westman (2005) who, using a Schr\"{o}dinger wave function $\psi$, demonstrate that an arbitrary initial choice for the ontological  or Bohm particle density $\rho$ relaxes to the Born probability density $|\psi|^2$, provided both densities are coarse--grained. The relaxation owes to the stirring action of the velocity field $ {\bi v}={\bi v}[\bi {x}, t]$ prescribed by de Broglie in terms of $\psi$:
\be
{\bi v} = {\bi j}/ |\psi|^2\; ,
\ee
where  ${\bi j}$ is the Schr\"{o}dinger current:
\be
{\bi j} =(\hbar /m) {\mathrm I}{\mathrm m} (\psi^*\nabla \psi)\; .
\ee
The demonstration by Valentini and Westman (2005) of the relaxation of the coarse--grained particle density to `quantum thermal equilibrium'  is a major development for the de Broglie--Bohm (`dBB') theory. Valentini (2008) has proposed that particles emitted in the early universe may not yet be in equilibrium. The analogy of the dBB formalism with fluid dynamics suggest borrowing the Lagrangian representation from the latter (e.g., Holland, 2009). It is shown here in \S2,  using the Lagrangian representation, that the relaxation demonstrated by the numerical computations of Valentini and Westman (2005) is in general to be expected. 

\section{Relaxation: Schr\"{o}dinger equation} 
\subsection{particle kinematics}
The Eulerian field ${\bi v}={\bi v}[{\bi x},t]$ of the de Broglie velocity (1) defines particle paths ${\bi x}={\bi P}({\bi a},s;t)$ through the Lagrangian prescription
\be
\frac{\partial {\bi P}} {\partial t}({\bi a},s;t)={\bi v}[{\bi P}({\bi a},s;t),t]\; ,
\ee
subject to the general initial condition
\be
{\bi P}({\bi a},s;s)={\bi a}\; .
\ee
Note that both $t>s$ and $t<s$ are allowed. In particular, if ${\bi x}={\bi P}({\bi a,}s;t)$  then ${\bi a}={\bi P}({\bi x}, t;s)$\;.
In the interest of clarity, Eulerian arguments are denoted by $[{\bi x}, t]$, Lagrangian arguments by $({\bi a}, s;t)$\;.
The Jacobian of the transformation ${\bi a} \to {\bi P}({\bi a}, s;t)$ is denoted by $J=J({\bi a},s; t)$  where
\be
J=\frac{\partial(P^1,P^2,P^3)}{\partial (a^1,a^2,a^3)}\;  .
\ee
In particular, $J({\bi a}, s;s)=1$\;. It is a kinematic identity (e.g., Lamb, 1932; Bennett, 2006) that 
\be
\frac{\partial J}{\partial t}({ \bi a}, s;t)= J({ \bi a}, s;t) \nabla\cdot {\bi v}[{\bi P}({\bi a},s;t),t]\; .
\ee
Arguments will not always be displayed with such care subsequently, but the meaning of the partial derivatives may be inferred from context. For any field $F[{\bi x}, t]$\;, the following relationship expresses the rate of change of $F$ following the motion both in Lagrangian and Eulerian terms:
\be
\frac{\partial F}{\partial t}({\bi a},s;t) = \frac{\partial F}{\partial t}[{\bi P}({\bi a}, s;t),t]+ {\bi v} [{\bi P}({\bi a}, s;t),t]\cdot \nabla F [{\bi P}({\bi a}, s;t),t]\; .
\ee

\subsection{Bohm particle density}
According to the dBB theory, the particle density $\rho = \rho [{\bi x}, t]$  obeys the Eulerian conservation law
\be
\frac{\partial \rho}{\partial t} + \nabla \cdot ({\bi v} \rho)=0\; .
\ee
This may be rearranged as
\be
\frac{\partial \rho}{\partial t} + {\bi v}\cdot \nabla \rho = -\rho \nabla \cdot {\bi v}\; ,
\ee
which by virtue of (6) and (7) has the Lagrangian form (Lamb, 1932, Art.14)
\be
\frac{\partial( \rho J)} {\partial t} ({\bi a}, s;t)=0\; .
\ee

\subsection{Born probability density}
All solutions $\psi=\psi[\bi{x}, t]$ of the Schr\"{o}dinger equation 
\be
\rmi \hbar \frac{\partial \psi}{\partial t} = -\frac{\hbar^2}{2m}\nabla^2\psi +V\psi\;,
\ee 
where V is the potential field (e.g., Feynman \etal,  1965), conform to the dynamical identity
\be
\frac{\partial |\psi|^2}{\partial t} +\nabla \cdot {\bi j}=0\; ,
\ee
where the Schr\"{o}dinger current ${\bi j}$ is defined in (2). This identity may be rearranged to resemble an Eulerian  `probability conservation law' involving the de Broglie velocity (1):
\be
\frac{\partial |\psi|^2}{\partial t} +\nabla \cdot({\bi v}|\psi|^2)=0\; ,
\ee
which has the Lagrangian form
\be
\frac{\partial( |\psi|^2 J)} {\partial t} ({\bi a}, s;t)=0\; .
\ee
It is clear from (10) and (14) that $\rho$ and $|\psi|^2$ are  identical at all times $t$ if and only if they are identical at one time $s$, say $s=0$\;. It is tacitly assumed that the particles are inside the domain $D$ for all time, which is assured if $\psi^*\nabla \psi$ is real on the boundary. This condition is satisfied by a variety of natural choices of boundary conditions for the Schr\"{o}dinger equation (11). It also suffices that $\psi$ be periodic in space (Valentini  and Westman, 2005).

\subsection{Coarse--graining}
The stirring of the particle density by relatively smooth fields of de Broglie velocity leads in general to very fine structure in the density field (Valentini and Westman, 2005). The structure may be suppressed by spatial smoothing or `coarse--graining' in space:
\be
\{\rho[{\bi x},t]\} =w^{-1}\int_{cell} \rho[{\bi y}, t] d\bi{y}\; , 
\ee
where the region of integration is a small but finite rectangular spatial cell centered on $\bi{x}$\;, the infinitesimal volume element  $d\bi{y}$ is $ dy^1dy^2dy^2$ and $w$ is the cell volume. Each point ${\bi y}$ in the cell at time $t$  lies on a path from a point ${\bi b}={\bi P}({\bi y}, t;0)$ somewhere inside $D$ at time  $t=0$\;, that is
\be
\{\rho[{\bi x},t]\} =w^{-1}\int_{{\cal S}} \rho({\bi b},0;t) J({\bi b}, 0;t)d{\bi b}\; , 
\ee 
where $\rho({\bi b}, 0;t)= \rho[{\bi P}({\bi b},0;t),t]$\;. The region of integration ${\cal S}={\cal S}({\bi x}, t)$ is the pre--image, at time $t=0$, of the cell centered on ${\bi x}$ at time $t$\;. In general ${\cal S}$ is a highly convoluted but simply connected region. Equation (9) shows that there is a tendency for particles to concentrate in zones of divergence ($\nabla \cdot {\bi v} >0)$ as time decreases, but the zones are in general so transient that the stirring spreads the cell  throughout $D$\;, albeit into a convoluted shape. By virtue of the conservation law (10), the coarse--grained density is
\be
\{\rho[{\bi x},t]\} =w^{-1}\int_{{\cal S}} \rho[{\bi b},0] d{\bi b}\; , 
\ee
since $ \rho({\bi b},0;0)=\rho[{\bi P}({\bi b}, 0;0), 0]=\rho[{\bi b}, 0]$\; , and $J({\bi b},0;0)=1$\;. Now at any time $t$ the particle density $\rho$ is normalized over  $D$, and in particular at $t=0$\;, thus
\be
\rho[{\bi b},0] =W^{-1}+\rho'[{\bi b},0]\; ,
\ee
where $W$ is the volume of $D$\;, and the integral of $\rho'$ over $D$ vanishes. Substituting for $\rho$ in the right hand side of (17) yields
\be
\{\rho[{\bi x},t]\} =(Ww)^{-1}\int_{{\cal S}}d{\bi b}+\ldots \, .
\ee
The ellipsis in (19) is small, owing to the integral over the `tentacles' of ${\cal S}$ sampling and summing the signed values of $\rho'$\;. The integral in (19) is the volume of ${\cal S}$, denoted ${\cal W}={\cal W}({\bi x}, t)$,  thus the the coarse--grained particle density is found to be 
\be
\{\rho[{\bi x},t]\} =(Ww)^{-1}{\cal W}[{\bi x},t]+\ldots=W^{-1}\{J({\bi x}, t;0)\} +\ldots\; .
\ee
By an identical argument it follows also that
\be
\{|\psi|^2[{\bi x},t]\} =(Ww)^{-1}{\cal W}[{\bi x},t]+\ldots=W^{-1}\{J({\bi x}, t;0)\} +\ldots\; .
\ee
In general,  $|\psi|^2$ is far less affected than $\rho$  by coarse--graining (Valentini and Westman, 2005). In any case (20) and (21) explain the relaxation of the coarse--grained Bohm particle density, irrespective of its initial values, to the coarse--grained Born probability density. They further explain why $|\psi|^2$ may be used as a probability density.

\section{Discussion}
The elementary manipulations leading to (20) and (21) are not found in, for example, Tolman (1938). The density considered there is that of particles in phase space, where the flow is solenoidal owing to the underlying classical dynamics being Hamiltonian. Indeed, if ${\bf x}$ in (20) represents a point in phase space, and if $J\equiv 1$\;, then $\rho[{\bf x },t]$ relaxes to the uniform density assumed by Tolman. The analysis in \S2 does not depend upon the dimensionality of ${\bf x}$\;, so relaxation also results from stirring by the current field of a Schr\"{o}dinger wave function on a higher--dimensional configuration space $({\bi x}_1, {\bi x}_2,\dots)$\;.  There are cases of interest in which relaxation does not occur, the most obvious being vanishing de Broglie velocity: ${\bi v}=0$, as in a standing Schr\"{o}dinger wave. A uniform but time dependent flow would preserve cell shape. Rigid rotation: ${\bi v}=\bOmega \times {\bi x}$, which would also preserve the cell shape, is not of interest since it is rotational: $\nabla \times {\bi v} = 2{\bi \bOmega}$\;. On the other hand a steady flow can stir effectively unless it restricts the cell particles to a region in which $\rho'$\;, the deviation of the initial particle density from its spatial mean, is one-signed.

The stirring efficiency of the de Broglie velocity field controls the relaxation. The velocity field (1), associated with the Schr\"{o}dinger wave function $\psi=|\psi|\exp(\rmi S/ \hbar)$\;, may also be expressed as ${\bi v} = m^{-1}\nabla S$\;. The dynamics of this `fluid' are revealed by decomposing the Schr\"{o}dinger equation (11) for the amplitude and phase of $\psi$ (e.g., Holland 1993).  The dynamics of $\nabla S $ resemble those of an inviscid fluid. However the field (1) is irrotational, and thus cannot be turbulent in the usual sense. The field (1) has divergence $\delta=\nabla \cdot {\bi v}$\;, and the `fluid dynamics' of this divergence may be rearranged as $\partial \delta/ \partial t + \delta^2 +....= ....\;,$ where the time derivative follows the motion.  The Riccati operator admits the possibility of blow--up in a finite time: $1/\delta(t) = 1/\delta(0)+t$\;. On the other hand $\psi$ is the solution of the linear Schr\"{o}dinger equation (11), and so the wavenumber spectrum  of $\psi$ is determined by the initial wave function and by the potential. Valentini and Westman (1995) report highly intricate and rapidly separating paths derived from a relatively smooth initial wave function and a uniform constant potential.  

Coarse--graining is crucial for relaxation of the Bohm particle density to quantum thermal equilibrium, and the formalism of a density at a single point is misleading. No matter how small the cell size, it will eventually (to speak in terms of time increasing) disperse into a space--exploring shape. A single particle remains forever a single particle. That is, relaxation is an effect of relative dispersion, not absolute. The relaxation time scale is not readily estimated in general. If the velocity field that stirs the particles is rich in fine structure then particles disperse independently, each taking a random walk, leading to root mean square particle separations growing like $\Or(t^{1/2})$\;. If the dominant stirring effect is on relatively large scales, then relative dispersion owes to the large--scale rate of strain and separations grow exponentially in time.  Random walks through simple shear flows are ubiquitous in the ocean and atmosphere, with particle separations growing like $\Or (t^{3/2})$\;. Random walks through a large--scale divergence lead to exponentially growing separations. In conclusion, the relaxation time scale cannot be expected to have a simple parameter dependence.

\section*{References}
\begin{harvard}

\item[]Bell J S 1987 {\it Speakable and Unspeakable in Quantum Mechanics} (Cambridge: Cambridge University Press) 

\item[] Bennett A 2006 {\it Lagrangian Fluid Dynamics (Cambridge Monographs on Mechanics)} (Cambridge: Cambridge University Press) 

\item[] Bohm D and Hiley  B J 1993 {\it The Undivided Universe: An Ontological Interpretation of Quantum Theory} 
(London: Routledge)

\item[] Feynman R P, Leighton R B and Sands M  1965 {\it The Feynman Lectures on Physics: Quantum Mechanics} (Reading, Addison--Wesley)

\item[] Lamb H 1932 {\it Hydrodynamics} (Cambridge: Cambridge University Press)

\item[] Holland P R 1993 {\it The Quantum Theory of Motion: an Account of the de Broglie--Bohm Causal Interpretation of Quantum Mechanics} (Cambridge: Cambridge University Press)

\item[]Holland P 2009 Schr\"{o}dinger dynamics as a two--phased conserved flow; an alternative trajectory construction of quantum propagation \JPA 42:075307 {\it Preprint} quant--ph/0807.4482

\item[] Passon O 2006 What you always wanted to know about Bohmian mechanics but were afraid to ask {\it Preprint}  quant--ph/0611032v1 

\item[] Tolman R C 1938 {\it The Principles of Statistical Mechanics} (Oxford, Clarendon Press)

\item[] Valentini A 2008  Inflationary cosmology as a probe of primordial quantum mechanics {\it Preprint} hep--th/0805.0163v1 

\item[] Valentini A and Westman H 2005 \PRS {\bf A461} 253

\item{}
\end{harvard}

\end{document}